\documentclass[tightenlines,aps,
twocolumn,
showpacs,nofootinbib]{revtex4}

\usepackage{psfig,float}
\usepackage{amsmath,amsfonts,graphicx,bm}

\newcommand{\be}{\begin{equation}}
\newcommand{\ee}{\end{equation}}
\newcommand{\ba}{\begin{eqnarray}}
\newcommand{\ea}{\end{eqnarray}}

\begin{document}

\begin{titlepage}

\begin{flushright}
\vbox{
\begin{tabular}{l}\\
 UH-511-1075-05\\
 MADPH-05-1436
\end{tabular}
}
\end{flushright}

\title{
The gluon-fusion uncertainty in Higgs coupling extractions
}

\author{
Charalampos Anastasiou\thanks{e-mail:babis@phys.ethz.ch}} 
\affiliation{
          Institute for Theoretical Physics,\\ 
          ETH, 8093  Z\"urich, Switzerland}
\author{Kirill Melnikov
        \thanks{e-mail: kirill@phys.hawaii.edu}}
\affiliation{Department of Physics and Astronomy,
          University of Hawaii,\\ 2505 Correa Rd., Honolulu, Hawaii 96822}  
\author{Frank Petriello\thanks{frankjp@physics.wisc.edu}}
\affiliation{
Department of Physics, Johns Hopkins University, \\
3400 North Charles St., Baltimore, MD 21218 \\ and \\ 
Department of Physics, University of Wisconsin, Madison, WI, 53706
} 

\begin{abstract}

We point out that the QCD corrections to the gluon-fusion Higgs boson production cross section at the LHC are very 
similar to the corrections to the Higgs decay rate into two gluons.  Consequently, the ratio of these two quantities has a 
theoretical uncertainty smaller than the uncertainty in the cross section alone by a factor of two.
We note that since this ratio is the theoretical input 
to analyses of Higgs coupling extractions 
at the LHC, the reduced uncertainty should be used; in previous studies, the full cross section uncertainty was employed.

\end{abstract}

\maketitle

\end{titlepage}

The benchmark for future new particle searches is the search for the Higgs boson at the LHC.  It is the only particle predicted 
by the Standard Model (SM) that has not been observed, and its discovery will be an important milestone for high energy physics.  
Experiments at LEP and SLC established a lower limit on the mass of the SM Higgs of $m_H > 114$ GeV through direct searches~\cite{Barate:2003sz}.  Indirect 
constraints from precision electroweak measurements indicate that a scalar Higgs boson with $m_H < 200$ GeV exists~\cite{:2004qh}.  This state will 
provide a powerful probe of the mechanism of electroweak symmetry breaking, which has motivated numerous analyses studying which of its properties can be 
measured: whether its $CP$ properties can be determined, if its production and decay modes can be accurately measured, and a host of other 
issues~\cite{Carena:2002es}.

A large effort has been devoted to studying the extraction of Higgs boson partial widths at both the LHC and a future linear collider~\cite{higgscoup}.  
The starting point for these analyses is the following relation for each 
Higgs production and decay channel:
\begin{equation}
\sigma_p(H) \times BR(H\rightarrow xx) = \frac{\sigma_p^{SM}(H)}{\Gamma_p^{SM}} \times \frac{\Gamma_p \Gamma_x}{\Gamma},
\label{mastereq}
\end{equation}
where $\Gamma_x$ denotes the partial width for the decay $H \rightarrow xx$, $\Gamma_p$ is the partial width for the Higgs production mode under 
consideration, 
and $\Gamma$ is the total Higgs width.
This equation is valid under two assumptions. First, the narrow-width 
approximation is used, which should hold up to intermediate Higgs boson 
masses. Second, it is assumed that a single, well-defined channel can be 
assigned to the Higgs production process.  The observed rate furnishes a measurement of the product $\Gamma_p \Gamma_x / \Gamma$, subject to experimental 
errors and the theoretical uncertainty in the normalization 
factor $\sigma_p^{SM}(H)/ \Gamma_p^{SM}$; the theoretical input and the experimental measurement are clearly separated using Eq.~\ref{mastereq}.  
The available production modes include gluon fusion, weak boson fusion, $t\bar{t}H$ associated 
production, and $WH$, $ZH$ associated production.  The accessible decay modes in each production channel are: $gg \rightarrow H \rightarrow WW,ZZ,
\gamma\gamma$; WBF with $H \rightarrow WW,ZZ,\gamma\gamma,\tau\tau$; $t\bar{t}H$ production with $H \rightarrow WW,\gamma\gamma,bb$; 
$WH$, $ZH$ associated production with $H \rightarrow WW,\gamma\gamma$.  
A combined analysis of the many channels available determines the various partial 
widths $\Gamma_x$.

The largest theoretical error entering this analysis is the uncertainty in the gluon-fusion production cross section, $gg \rightarrow H$.  
Next-to-leading 
order (NLO) QCD calculations of this cross section revealed corrections reaching nearly 70\% and exhibiting large residual scale 
dependences~\cite{higgsnlo}.  
This stimulated the 
calculation of the NNLO corrections.  The inclusive NNLO result for $gg\rightarrow H$ was obtained in~\cite{higgsnnlo}; very recently the leading N$^3$LO 
corrections to the inclusive cross-section  were computed \cite{moch}.
The completely differential Higgs 
production cross section, which allows a realistic study of experimental acceptances at NNLO, was calculated in~\cite{fehip}.  The partial width 
$\Gamma_{gg}^{SM}$ was also calculated to NNLO~\cite{Chetyrkin:1997iv}.  However, in spite of this significant effort, the $gg \rightarrow H$ channel
still suffers from large theoretical uncertainties.  The residual scale dependences in both the cross section and partial width are at the 
20\% level.  In the most recent study of Higgs couplings at the LHC~\cite{Duhrssen:2004cv}, a $\pm 20\% $ theoretical error was 
assigned to the $gg \rightarrow H$ channel. 
The next largest 
error entering the analysis is a 15\% uncertainty in the $t\bar{t}H$ production cross section; the remaining theoretical errors are under 10\%, as are the 
individual experimental systematics.

\noindent
\begin{figure}[htb]
\centerline{
\psfig{figure=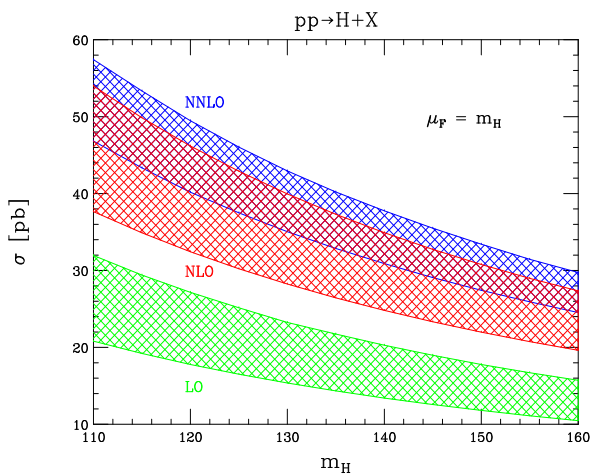,height=6.2cm,width=7.8cm,angle=90}}
\vspace{-0.2cm}
\caption{$\mu_R$ scale dependence for the Higgs production cross section at the LHC, as a function of $m_H$.  $\mu_R$ is varied 
between $m_h/2 \leq \mu_R \leq 2m_h$, while $\mu_F = m_H$.  The LO, NLO, and NNLO distributions are shown.}
\label{sigmaR}
\end{figure}
%

In this  note, we point out that the appropriate uncertainty in the 
$gg\rightarrow H$ channel which enters the analysis of Higgs couplings should 
instead be $\pm 5\%$, which is smaller by a factor of four.  This reduction 
relies upon the observation that the theoretical input 
for the Higgs coupling 
determination is the ratio
$\sigma_{gg}^{SM} / \Gamma_{gg}^{SM}$.  
The QCD corrections to $\sigma_{gg}^{SM}$ and $\Gamma_{gg}^{SM}$ 
track each other, and a large portion of the uncertainty cancels 
when the ratio is taken.  We believe that the $\pm 20\%$ error 
assumed in the analysis of \cite{Duhrssen:2004cv} is overly conservative.  The NNLO 
Higgs production itself suffers from an uncertainty of only $\pm 10\%$~\cite{higgsnnlo}, and as we will 
argue later, the imprecise knowledge of the gluon 
distribution function is smaller than the residual scale uncertainty.
Since this is the 
largest single error entering the analysis of Higgs couplings, its reduction 
may  have an important effect on the precisions of the $\Gamma_x$.


We now briefly explain why the theoretical uncertainty 
in the ratio is reduced, and then present a  numerical proof.  The perturbative expansions for the cross section and partial width have the forms
\begin{eqnarray}
\sigma_{gg}^{SM} &\sim& \alpha_s^2(\mu_R) C^2 \left\{ 1+\alpha_s(\mu_R)X_1(\mu_R,\mu_F) \right. \nonumber \\ 
          &+& \left. \alpha_s^2(\mu_R)X_2(\mu_R,\mu_F) +\ldots \right\}, \nonumber \\
\Gamma_{gg}^{SM} &\sim& \alpha_s^2(\mu_R) C^2 \left\{ 1+\alpha_s(\mu_R)Y_1(\mu_R) \right. \nonumber \\
          & +& \left. \alpha_s^2(\mu_R)Y_2(\mu_R) +\ldots \right\}.
\label{pexp}
\end{eqnarray}
$\mu_R$ and $\mu_F$ respectively denote the renormalization and factorization scales, while $C$ denotes the Wilson coefficient obtained 
from integrating out the top quark.  The width has no $\mu_F$ dependence, as no initial-state mass-factorization is required.  An immediate 
consequence of these formulae is that both $\sigma_{gg}^{SM}$ and $\Gamma_{gg}^{SM}$ are proportional to the same factor of 
$\alpha_s^2(\mu_R) C^2$.  This quantity 
contributes  a large renormalization scale dependence to both $\sigma_{gg}^{SM}$
and $\Gamma_{gg}^{SM}$. 
The same scale $\mu_R$ can be chosen in both calculations, which completely removes the large renormalization scale uncertainty when the ratio 
$\sigma_{gg}^{SM} / \Gamma_{gg}^{SM}$ is taken.  When integrating out the top quark to obtain the Wilson coefficient $C$, the relevant scale is 
the top quark mass $m_t$; this choice removes all large logarithms from the expression.  This dependence is identical in both $\sigma_{gg}^{SM}$ and 
$\Gamma_{gg}^{SM}$, and therefore cancels exactly.
The observation that $\mu_R$ is approximately the same for both
$\sigma_{gg}^{SM}$ and $\Gamma_{gg}^{SM}$ in the remaining pieces relies upon the fact that
the Higgs boson production  is dominated by partonic threshold.  This implies that the partonic center-of-mass energy squared $\hat{s}$,
relevant for the $gg \to H$ process,
is $\hat s \sim m_{H}^2$.  Since $ m_H$ is also the only mass scale
that enters the decay rate, the kinematic
scales for the production and decay processes must be similar.  This indicates that identical $\mu_R$ should be chosen for both processes, 
and also implies that the $\mu_R$ dependence of the expansion coefficients for 
the cross section and width should be similar, $X_i \sim Y_i$.  This further reduces the renormalization scale dependence.
Since the factorization scale uncertainty turns out to be small,
the cancellation of the renormalization
scale dependence leads to  a reduced theoretical error in the ratio.

\begin{figure}[htb]
\centerline{
\psfig{figure=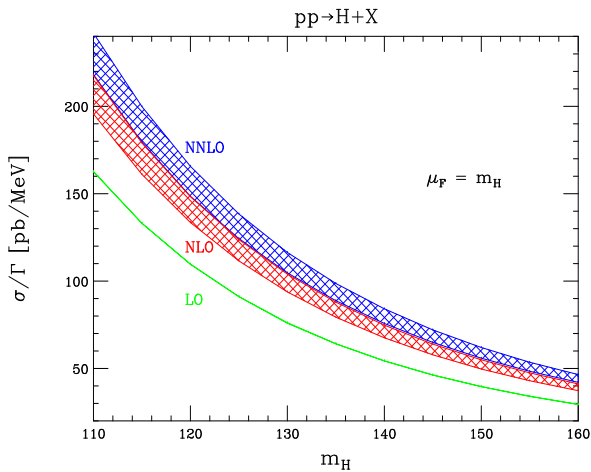,height=6.2cm,width=7.8cm,angle=90}}
\vspace{-0.2cm}
\caption{$\mu_R$ scale dependence for the $\sigma / \Gamma$ ratio at the LHC, as a function of $m_H$.  $\mu_R$ is varied
between $m_h/2 \leq \mu_R \leq 2m_h$, while $\mu_F = m_H$.  The LO, NLO, and NNLO distributions are shown.}
\label{ratioR}
\end{figure}

We now present numerical proof of this assertion.
We estimate the theoretical errors by varying both scales within 
the range $m_H/2 \leq \mu_R,\mu_F \leq 2m_H$.
We vary each scale in the range $m_H/2 \leq \mu_R,\mu_F \leq 2m_H$, and fix the other scale to $m_H$.  The $\mu_R$ variation of the 
cross section is 
20\% to 25\% even at NNLO (see Fig.~\ref{sigmaR}), 
while the $\mu_F$ variation is less than 5\% at NNLO and is therefore negligible compared with $\mu_R$ scale variation.  
The $\mu_R$ dependence is reduced to the 10\% level when the ratio is
taken, as shown in Fig.~\ref{ratioR}, while the $\mu_F$ dependence 
is unchanged.  We note that because of the complete cancellation of the $\mu_R$ dependence at LO, we do not obtain a reliable 
estimate of the theoreical uncertainty until the NNLO corrections are included.
We use the program \textsf{FEHiP}~\cite{fehip} to obtain
cross section numbers, and we take the width results from~\cite{Chetyrkin:1997iv}.  We believe that the scale variation bands presented 
here are a good estimate of the theoretical uncertainty: the NLO and NNLO bands overlap, the fixed order results are in good agreement 
with those obtained with threshold resummation~\cite{Catani:2003zt}, and the leading ${\rm N^3LO}$ corrections do not significantly 
change the NNLO results~\cite{moch}.
\noindent

\begin{figure}[htb]
\centerline{
\psfig{figure=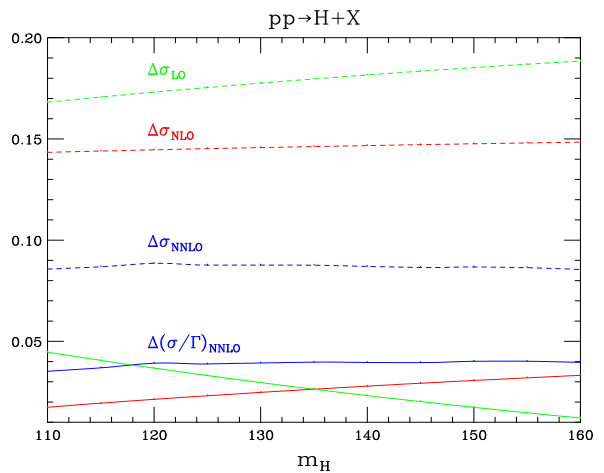,height=6.2cm,width=7.8cm,angle=90}}
\vspace{-0.2cm}
\caption{Theoretical error on $\sigma$ and $\sigma/\Gamma$ as a function of Higgs mass.  The LO, NLO, and NNLO
uncertainties are shown.  The green and red lines at the bottom denote the theoretical errors at LO and NLO, respectively.  As 
explained in the text, only at NNLO is a reliable error estimate dependence obtained.}
\label{uncert}
\end{figure}

In Fig.~\ref{uncert}, we estimate the remaining theoretical error in 
${\cal O}=\sigma_{gg}^{SM},\sigma_{gg}^{SM}/\Gamma_{gg}^{SM}$ 
by first setting $\mu_R=\mu_F=\mu$, 
and defining $\Delta {\cal O} = |{\cal O}(\mu=2m_H)-{\cal O}(\mu=m_H/2)|/2{\cal O}_{\rm average}$.  We can then find the appropriate error 
band by taking $\pm \Delta {\cal O}$.
In both cases there is a slight cancellation between the $\mu_R$ and 
$\mu_F$ dependences;  we conservatively round the error up and estimate $\Delta \sigma \approx \pm 10\%$ and 
$\Delta (\sigma/\Gamma) \approx \pm 5\%$.  As claimed, a large portion of the theoretical uncertainty cancels in the ratio.

We also point out that the cancellation of the QCD uncertainties between 
$\sigma_{gg}$ and $\Gamma_{gg}$ is expected even after mild cuts 
on the final states are imposed. For example, the calculation of~\cite{fehip} shows that the QCD corrections to 
$\sigma(gg \to H \to \gamma \gamma)$ are quite similar to that of the 
total $gg \to H$ cross section even if the standard cuts used by ATLAS 
and CMS for the identification of the two-photon signal 
\cite{cuts} are applied. 
We also  
expect a reduction of the theoretical 
uncertainty in quantities of the type 
$\sigma(gg \to H +~{\rm jets})/\Gamma_{gg}$. However, this reduction should 
be less effective than in the total cross section because 
$\sigma(H + {\rm jets})$ depends on additional kinematic factors, 
such as the $p_\perp$ of the jets, that control the magnitude 
and scale dependence of QCD effects.
 
Finally, we note that the Higgs boson production cross section suffers 
from imprecise knowledge of  parton distribution functions (pdfs). 
The resulting uncertainties strongly depend on the production mode and the 
mass of the Higgs boson; they were recently analyzed in 
\cite{Djouadi:2003jg}. For $gg \to H$ and a Higgs boson with mass 
$m_H \sim 200~{\rm GeV}$, the pdf errors lead to an uncertainty 
in the Higgs boson production cross section of about $\pm 2.5\%$, 
smaller than the scale variations studied here and negligible when added in quadrature.
In contrast to the renormalization and factorization scale 
uncertainties, the pdf error is {\it reducible}, 
since measurements of Standard Model processes at the LHC 
may lead to more accurate pdf determinations. For these reasons, we do not 
consider the pdf uncertainty in what follows.

The  $\pm 20\%$ uncertainty on the gluon fusion production cross section 
is the dominant uncertainty in the analysis of \cite{Duhrssen:2004cv} 
for $m_H > 150~{\rm GeV}$.  While the full re-analysis is  required 
to access the impact of the uncertainty reduction that we  advocate, 
some simple estimates can be made. Consider the 
determination of $\Gamma_{gg}$ from the 
production of $W$-bosons in the gluon fusion, $gg \to H \to W^+W^-$. 
The uncertainty in 
$\Gamma_{gg}$ is given by 
\begin{equation}
\frac{\delta \Gamma_{gg}}{\Gamma_{gg}} =
\sqrt{ \left ( \frac{\delta N}{N} \right )^2 
+ \left ( \frac{\delta T}{T} \right )^2
+ \left ( \frac{\delta \Gamma}{ \Gamma} \right )^2 
+ \left ( \frac{\delta \Gamma_W }{ \Gamma_W} \right )^2 },
\end{equation}
where $N$ is the number of events, $T = \sigma_{gg}^{SM}/\Gamma_{gg}^{SM}$ 
is the 
theoretical normalization factor and $\Gamma_W$ is the decay rate for 
$H \to WW$. For the purpose of this estimate we assume that 
the errors are not correlated.
Following \cite{Duhrssen:2004cv}, we use $\delta N/N \approx 10^{-1}$, 
$\delta \Gamma/\Gamma \approx 1.4 \times 10^{-1}$ and 
$\delta \Gamma_W/\Gamma_W 
\approx 10^{-1}$, for $m_H \approx 160~{\rm GeV}$.  Using $\delta T/T 
\approx 2 \times 10^{-1}$, we find that the uncertainty on $\Gamma_{gg}$ 
is thirty percent.  As stressed here, it is more appropriate
to use $\delta T/T \approx 5 \times 10^{-2}$; the uncertainty on  
$\Gamma_{gg}$ then becomes about twenty percent.  
An analysis is underway to determine the effect of the error reduction 
on the extraction of Higgs properties at the LHC~\cite{dpc}.

\noindent


We thank T. Plehn and D. Rainwater for many useful discussions.
The work of K. M. was supported by the US Department of Energy  under contract DE-FG03-94ER-40833 and  
the Outstanding Junior Investigator Award DE-FG03-94ER-40833, and by the 
Alfred P. Sloan Foundation.  The work of F. P. was supported
by the National Science Foundation  under 
contracts P420D3620414350, P420D3620434350, and by the Univeristy of Wisconsin Research Committee with funds 
granted by the Wisconsin Alumni Research Foundation.


\end{document}